\documentclass[12pt]{article}

\catcode`\@=11
\@addtoreset{equation}{section}

\evensidemargin \oddsidemargin

\usepackage{mathrsfs,amsbsy,amssymb,latexsym,amsfonts,amsmath,
amssymb,
fancyhdr,lastpage,
fancybox,
ulem,cases
,wrapfig}

\usepackage[dvipdfmx]{graphicx}

\usepackage{pifont}

\usepackage{color}
  \usepackage{ascmac}
  
  \usepackage{bm}

\newcommand\CW{\mathcal{W}}

\newcommand\ab{\boldsymbol{a}}

\newcommand\bb{\boldsymbol{b}}

\newcommand\pa{\partial}

\newcommand\phib{\boldsymbol{\phi}}
\newcommand\pacb{{\pa\boldsymbol{c}}}
\newcommand\xib{\boldsymbol{\xi}}

\newcommand\intd{\int\!\d^Dx\,}

\newcommand\nn{\nonumber}

\newcommand\disp{\displaystyle}

\newcommand\phis{{\phi^*}}
\newcommand\phibs{{\phib^*}}

\newcommand\tr{{\rm tr}\,}

\renewcommand\d{{\rm d}}

\newcommand\bref[1]{(\ref{#1})}

\newcommand\pat{\tilde\pa}
\newcommand\Gt{\tilde G}

\begin{document}

\vspace*{3cm}

\begin{center}
{\Large \bf 
On
Interacting 
Higher-Spin 
Bosonic Gauge Fields
\bigskip

in BRST-antifield Formalism
}
\vspace*{3cm}\\
Makoto Sakaguchi\footnote{\texttt{makoto.sakaguchi.phys@vc.ibaraki.ac.jp}}
and
Haruya Suzuki\footnote{\texttt{h.suzuki.ibaraki@gmail.com
}}

\end{center}
\vspace*{1.0cm}

\begin{center}

Department of Physics, Ibaraki University, Mito 310-8512, Japan
\end{center}

\vspace{2cm}

\begin{abstract}

We examine interacting bosonic higher-spin gauge fields
in the BRST-antifield formalism.
Assuming that an interacting action $S$ is
a deformation of the free action
with a deformation parameter $g$,
we solve the master equation $(S,S)=0$
from the lower orders in $g$.
It is shown that,
choosing
a certain cubic interaction
as the first-order deformation,
we can solve the master equation
and obtain an action containing all orders in $g$.
The anti-ghost number of
the obtained action is  
less than or equal to two.
Furthermore we
show that the obtained  action
is lifted to
that of interacting bosonic higher-spin gauge fields on anti-de Sitter spaces.

\end{abstract}

\thispagestyle{empty}
\setcounter{page}{0}

\newpage

\setcounter{footnote}{0}

\section{Introduction}

Higher spin gauge theories have been studied
since the 1930s
from various viewpoints.
For example,
they are expected to reveal characteristic aspects of string theory in the high-energy limit.
String theory may be regarded as a spontaneous symmetry-breaking phase
of higher-spin gauge theories \cite{Gross}.
Free higher-spin gauge theories are now well understood.
There are obstacles to constructing
consistent interactions,
one of which is 
the no-go theorem \cite{no-go}\footnote{
In \cite{KU81}, the properties of bosonic and fermionic particles with spin $j\geq 1$
are examined
in  the field theory framework.
}.
To avoid this,
the number of derivatives
contained in interaction vertices
should be restricted
so that  higher-spin gauge fields are not included in the asymptotic states.
For cubic vertices, the allowed number of derivatives is clarified by using a light-cone formulation
in \cite{Metsaev lc cubic}\cite{Metsaev lc cubic 2}.
For bosonic gauge fields,
vertices are constructed explicitly by using Noether's procedure in
\cite{BBvD85}\cite{MMR Noether}.
In constructing vertices,
a generalized curvature tensor \cite{dWitFreedman}
is frequently used as a building block.
It is gauge invariant and defined as the $s$-curl of a totally symmetric spin-$s$ bosonic gauge field.
It is not an easy task to construct a gauge-invariant  generalized curvature tensor
 on  general backgrounds
including anti-de Sitter (AdS) spaces.
In contrast, we will employ the Fronsdal tensor
\cite{Fronsdal}
as a building block,
and look for vertices on AdS spaces.
It is also difficult to construct
full interactions including not only cubic
but also higher-order vertices.
We usually construct an interacting theory as a small deformation of the free theory
including a few lower  orders in the deformation parameter $g$.
It may not be consistent
to construct vertices beyond the cubic order.
Even if we could construct interactions to any order,
the full interacting action may not be written in a closed form.

\medskip

In this paper
we will construct actions of interacting bosonic higher-spin gauge fields
on $D$-dimensional spacetimes
in the BRST-antifield formalism.
The BRST-antifield formalism is known to be very powerful in
constructing interactions systematically \cite{BRST-antifield}.
Employing this cohomological method,
interaction terms are constructed systematically in \cite{BRST cohomology}.
In the present paper, we will first construct actions on a flat spacetime using this method,
and then lift them to those on AdS spaces.
For this we will use the Fronsdal tensor as a building block.
To avoid the no-go theorem, spins of gauge fields are restricted appropriately.
We will comment on this point in the last section.
In section \ref{sec:two fields}, two gauge fields of spins $s$ and $2s$
are examined,
and
three gauge fields of spins $s_1$, $s_2$, and $(s_1+s_2)$
are examined in section \ref{sec:three fields}.
In either case,
we construct a full action
$S$ as a deformation of the free action $S^0$,
$S=S^0+gS^1+g^2S^2+\cdots$.
It is shown that
choosing  a certain cubic vertex
as $S^1$,
the master equation $(S,S)=0$ can be solved
from the lower order in $g$.
It is worth noting that the obtained action $S$ contains all orders in $g$
and
is written in a closed form by using a geometric series.
The action can  be rearranged as $S=S_0+S_1+S_2$ where the anti-ghost number of $S_a$ is $a$.
We extract the gauge symmetry
of the interacting action $S_0$
from the BRST symmetry $\delta_B^g X=(X,S)$.
Furthermore, we will write down 
actions of interacting bosonic higher-spin gauge fields on AdS spaces
from those on a flat spacetime.

\medskip

The paper is organized as follows.
In the next section, 
clarifying our notations we present the free action in the BRST-antifield formalism.
The BRST deformation scheme is explained
in section 3.
In section 4,
two interacting gauge fields
with spins $s$ and $2s$
are examined.
The master equation is solved from the lower orders in $g$
and the full action containing all orders in $g$ is derived.
In section 5,
the full action for 
three interacting gauge fields
with spins  $s_1$, $s_2$, and $(s_1+s_2)$
is derived
(a derivation of low-order deformations is given in appendix B).
In section 6
we write down actions of higher-spin bosonic gauge fields
on AdS spaces.
The last section is devoted to a summary and discussions.
In appendix A, we explain how a $\Gamma$-exact term results in a BRST-exact term.

\section{Free higher-spin bosonic gauge theory}  

We explain the notations used in this paper
and present the free action in the BRST-antifield formalism.

We introduce a totally symmetric rank-$s$ bosonic tensor 
field $\phi_{\mu_1\cdots\mu_s}$
in  a $D$-dimensional flat spacetime.
The Fronsdal tensor
\cite{Fronsdal} for $\phi_{\mu_1\cdots\mu_s}$
is defined by\footnote{
Our convention for symmetrization of indices is
\begin{align}
\pa_{(\mu_1}\cdots\pa_{\mu_r}\phi_{\mu_{r+1}\cdots\mu_n)\nu_1\cdots\nu_m}
=\frac{1}{r! (n-r)!}\sum_{\{\mu_1,\ldots,\mu_n\}}
\pa_{\mu_1}\cdots\pa_{\mu_r}\phi_{\mu_{r+1}\cdots\mu_n\nu_1\cdots\nu_m}
\,,
\nn
\end{align}
where $\{\mu_1,\ldots,\mu_n\}$ indicates that 
the sum is taken over permutations of $\mu_1,\ldots,\mu_n$.
}
\begin{align}
F_{\mu_1\cdots\mu_s}(\phi)
=
\square\phi_{\mu_1\cdots\mu_s}
-\pa_{(\mu_1} \pa\cdot\phi_{\mu_2\cdots\mu_s)}
+\pa_{(\mu_1}\pa_{\mu_2}\phi'_{\mu_3\cdots\mu_s)}~,
\label{Fronsdal flat}
\end{align}
where $\square
=\eta^{\mu\nu}\pa_\mu\pa_\nu$
with $\eta_{\mu\nu}=\mathrm{diag}(-1,+1,\ldots,+1)$.
A prime on $\phi$ represents the trace of $\phi$, 
namely $\phi'_{\mu_3\cdots \mu_s}=\phi_{\mu_3\cdots \mu_s}{}^\rho{}_\rho$\,,
and similarly
$\phi''_{\mu_5\cdots \mu_s}=\phi_{\mu_5\cdots \mu_s}{}^\rho{}_\rho{}^\sigma{}_\sigma$\,.
The divergence of $\phi$
is expressed as
 $\pa\cdot\phi_{\mu_2\cdots\mu_s}=\pa^{\mu_1}\phi_{\mu_1\mu_2\cdots\mu_s}$\,.

The Fronsdal equation $F_{\mu_1\cdots\mu_s}=0$ 
means that
when $s=1$ the Maxwell equation for the gauge field $A_\mu$ is
 $\square A_\mu-\pa_\mu \pa^\nu A_\nu=0$,
while when $s=2$
the linearized Einstein equation for the graviton $h_{\mu\nu}$ is
$\square h_{\mu\nu}-\pa_{(\mu} \pa^\rho h_{\nu)\rho}-\pa_\mu\pa_\nu h^\rho{}_\rho=0$\,.
The gauge transformation of $\phi_{\mu_1\cdots\mu_s}$,
which is a natural generalization of $\delta A_\mu=\pa_\mu \xi$ for $s=1$
and $\delta h_{\mu\nu}=\pa_{(\mu}\xi_{\nu)}$  for $s=2$,
is given by
\begin{align}
\delta \phi_{\mu_1\cdots\mu_s}=\pa_{(\mu_1}\xi_{\mu_2\cdots\mu_s)}~,
\end{align}
where $\xi$ is the rank-$(s-1)$ gauge parameter.
It is straightforward to see that
\begin{align}
\delta F_{\mu_1\cdots\mu_s}(\phi)=3\pa_{(\mu_1}\pa_{\mu_2}\pa_{\mu_3}\xi'_{\mu_4\cdots\mu_s)}~.
\label{eqn:delta F flat}
\end{align}
Here and hereafter,
we require that the gauge parameter is traceless for $s\geq 3$
\begin{align}
\xi'_{\mu_4\cdots\mu_s}=0~,
\end{align}
which implies the gauge invariance of 
the Fronsdal tensor.
Furthermore we
impose
a double traceless constraint on $\phi$
\begin{align}
\phi''_{\mu_5\cdots\mu_s}=0~
\label{psi''=0}
\end{align}
for $s\geq 4$,
so that the Fronsdal equation should describe
the propagation of a massless spin-$s$ gauge field.
The action which leads to the Fronsdal equation is given by
\cite{dWitFreedman}
\begin{align}
S_\mathrm{dWF}
&=\frac{1}{2}\int \d ^Dx\, \phi^{\mu_1\cdots\mu_s}G_{\mu_1\cdots\mu_s}(\phi)~.
\label{action phi}
\end{align}
We have introduced $G_{\mu_1\cdots\mu_s}$ as
\begin{align}
G_{\mu_1\cdots\mu_s}(\phi)&\equiv
F_{\mu_1\cdots\mu_s}
-\frac{1}{2}\eta_{(\mu_1\mu_2}F'_{\mu_3\cdots\mu_s)}
~,
\label{eqn:G}
\end{align}
where $F$ is the Fronsdal tensor defined in \bref{Fronsdal flat}.
A useful and important relation we frequently use in this paper is
\begin{align}
&\int\d^D x\, \varphi^{\mu_1\cdots\mu_s} \Gt_{\mu_1\cdots\mu_s}(\psi)
=\int\d^D x\, 
\Gt_{\mu_1\cdots\mu_s}(\varphi)
\psi^{\mu_1\cdots\mu_s} 
\,,
\label{eqn:Gt}
\end{align}
where $\psi$ and $\varphi$ are arbitrary totally symmetric rank-$s$ fields.
We have defined $\Gt(A)$ by
\begin{align}
\Gt_{\mu_1\cdots\mu_s}(A)\equiv&\,
G_{\mu_1\cdots\mu_s}(A)
+\frac{1}{2}\eta_{(\mu_1\mu_2}\pa_{\mu_3}\pa_{\mu_4}A''_{\mu_5\cdots\mu_s)}
\,,
\label{eqn:Gt def}
\end{align}
where $A$ is an arbitrary totally symmetric rank-$s$ field.
When the double traceless conditions on $\psi$ and $\varphi$
would be implemented, 
one obtains 
$\Gt(\psi)=G(\psi)$ and $\Gt(\varphi)=G(\varphi)$,
so that
$\int\d^D x\, \varphi^{\mu_1\cdots\mu_s} G_{\mu_1\cdots\mu_s}(\psi)
=\int\d^D x\, 
G_{\mu_1\cdots\mu_s}(\varphi)
\psi^{\mu_1\cdots\mu_s} 
$\,.
It follows  that, varying the action $S_\mathrm{dWF}$ with respect to 
$\phi_{\mu_1\cdots\mu_s}$,
we obtain 
the equation of motion $G_{\mu_1\cdots\mu_s}(\phi)=0$\,.
Taking the trace of
$G_{\mu_1\cdots\mu_s}(\phi)=0$\,,
we obtain $F'_{\mu_3\cdots\mu_s}=0$
by using $F''_{\mu_5\cdots\mu_s}=0$, which follows from $\phi''=0$.
As a result, the equation of motion implies the Fronsdal equation $F_{\mu_1\cdots\mu_s}=0$\,.

\subsection{Free action in the BRST-antifield formalism}
Corresponding to the gauge parameter $\xi_{\mu_1\cdots\mu_{s-1}}$,
we introduce a rank-$(s-1)$ Grassmann-odd ghost field $c_{\mu_1\cdots\mu_{s-1}}$
with the same algebraic symmetry.
The $c_{\mu_1\cdots\mu_{s-1}}$ must be traceless $c'_{\mu_3\cdots\mu_{s-1}}=0$,
just like $\xi'_{\mu_3\cdots\mu_{s-1}}=0$.
The gauge invariance of $S_\mathrm{dWF}$
in \bref{action phi} is encoded to
the BRST invariance under the BRST transformations
\begin{align}
\delta_B \phi_{\mu_1\cdots\mu_s}&=\pa_{(\mu_1}c_{\mu_2\cdots\mu_s)}~,~~~
\label{BRST phi}\\
\delta_B c_{\mu_2\cdots\mu_s}&=0~.
\label{BRST c}
\end{align}
The gauge field $\phi$ and the ghost field $c$ are collectively called ``fields'' and denoted as
 $\Phi^A
 $. 
\begin{wraptable}[10]{r}{6.5cm}\vspace{0mm}
\begin{align*}
\begin{array}{|c|c|c|c|}\hline 
Z & pgh(Z) & agh(Z) & gh(Z) \\\hline 
\phi_{\mu_1\cdots\mu_s} & 0 & 0 & 0 \\\hline 
c_{\mu_1\cdots\mu_{s-1}} & 1 & 0 & 1 \\\hline 
\phi^*_{\mu_1\cdots\mu_s} & 0 & 1 & -1 \\\hline 
c^*_{\mu_1\cdots\mu_{s-1}} & 0 & 2 & -2 \\\hline 
\end{array}
\end{align*}
   \caption{Grading properties of fields and antifields}
  \label{table:charges}
\end{wraptable}
We further introduce ``antifields'' $\Phi^*
_A=\{\phi^*_{\mu_1\cdots\mu_s},~c^*_{\mu_1\cdots\mu_{s-1}}\}$
which have the same algebraic symmetries but opposite Grassmann parity.
Two gradings are introduced.
One is the pure ghost number $pgh$, and the other is the antighost number $agh$.
The ghost number $gh$ is defined as $gh\equiv pgh-agh$\,.
The grading properties are summarized in Table.\ref{table:charges}.
The antibracket
for two functionals, $X(\Phi^A,\Phi^*_A)$ and $Y(\Phi^A,\Phi^*_A)$,
is defined by
\begin{align}
(X,Y)\equiv 
X\frac{\stackrel{\leftarrow}{\delta}}{\delta\Phi^A}\frac{\stackrel{\rightarrow}{\delta}}{\delta\Phi^*_A}Y
-X\frac{\stackrel{\leftarrow}{\delta}}{\delta\Phi^*_A}\frac{\stackrel{\rightarrow}{\delta}}{\delta\Phi^A}Y~.
\end{align}
The action $S_\mathrm{dWF}$
in \bref{action phi}
can be extended to $S^0[\Phi,\Phi^*]$ such that
the BRST transformation of a functional $X(\Phi^A,\Phi^*_A)$
is expressed as
\begin{align}
\delta_B X=(X,S^0)
\,.
\end{align}
Note that $\delta_B$ acts from the right.
The nilpotency $\delta_B^2=0$
requires the master equation $(S^0,S^0)
=0$\,.
In the present case,  the free action is found to be
\begin{align}
S^0[\Phi,\Phi^*]=\int \d ^Dx\, \left(
\frac{1}{2}\phi^{\mu_1\cdots\mu_s}G_{\mu_1\cdots\mu_s}(\phi)
+\phi^*{}^{\mu_1\cdots\mu_s}
\pa_{(\mu_1}c_{\mu_2\cdots\mu_s)}
\right)~,
\label{eqn:S^0}
\end{align}
which leads to
\bref{BRST phi}
and
\bref{BRST c},
and
\begin{align}
\delta_B\phi^*_{\mu_1\cdots\mu_s}&=-G_{\mu_1\cdots\mu_s}(\phi)~,~~~
\label{BRST phi*}\\
\delta_B c^*_{\mu_2\cdots\mu_s}&=
-s\pa\cdot\phi^*_{\mu_2\cdots\mu_s}
+
\frac{s}{D+2s-6}
\eta_{(\mu_2\mu_3}\pa\cdot\phi^*{}'_{\mu_4\cdots\mu_{s})}
~.
\label{BRST c*}
\end{align}
Note that the second term on the right-hand side of \bref{BRST c*}
is required for the nilpotency $\delta_B^2 c^*=0$.
It is straightforward to see that
this term
leaves
the action \bref{eqn:S^0}
unchanged.

\section{BRST deformation}\label{sec:BRST deformation}

The BRST deformation scheme is very useful in constructing interactions systematically
\cite{BRST-antifield}.
We will explain relevant aspects in this section.

Suppose that
$S$ is a deformation of $S^0$ expanded in a deformation parameter $g$,
\begin{align}
S=S^0+gS^1+g^2S^2+\cdots~.
\end{align}
When $S$ solves the master equation $(S,S)=0$,
$S$ is invariant under the BRST transformation
generated by $S$: $\delta_B^g
 S\equiv (S,S)=0$.
 Here we add the letter $g$ to $\delta_B$
 in order to distinguish it from $\delta_B$ generated by $S^0$
 considered in the preceding section.
The master equation $(S,S)=0$
means, at the order of $g^n$,
\begin{align}
\sum_{k=0}^n(S^k,S^{n-k})=0~.
\label{eqn:master equation g^n}
\end{align}
The equation for $n=0$, $(S^0,S^0)=0$,
is satisfied by definition.
It is worth noting that
$S^n$ is determined
by 
all of
the lower order terms,
$S^k$ ($k<n$).
To obtain higher order terms,
we must start examining
$S^1$ first of all.

\medskip

$S^1$ is determined by the equation \bref{eqn:master equation g^n}
for $n=1$,
\begin{align}
(S^1,S^0)&=0~,
\label{ME n=1}
\end{align}
which
means that $S^1$ is BRST invariant,\footnote{
Let $\ab$ be a $D$-form defined by $S^1=\int \ab$;
this is equivalent to 
$
\delta_B \ab + \d \bb=0
$
\,.
Since a BRST-exact part of $\ab$ corresponds to a trivial field redefinition,
we consider $H^0(\delta_B|\d)$: the cohomology of the BRST differential $\delta_B$ modulo $\d$
at $gh=0$.
}
\begin{align}
\delta_B S^1=0\,.
\label{BRST closed}
\end{align}
If $S^1$ is BRST exact, 
namely $S^1=(X,S^0)
=X\frac{\stackrel{\leftarrow}{\delta}}{\delta\Phi^A}\frac{\stackrel{\rightarrow}{\delta}S^0}{\delta\Phi^*_A}
-X\frac{\stackrel{\leftarrow}{\delta}}{\delta\Phi^*_A}\frac{\stackrel{\rightarrow}{\delta}S^0}{\delta\Phi^A}
$~,
$S^1$ can be eliminated by a field redefinition.
In fact, 
the field redefinition $\Phi^A\to\Phi^A + g u^A$
and $\Phi^*_A\to\Phi^*_A + g v_A$
causes
the change
$S^0[\Phi^A,\Phi^*_A]\to S^0[\Phi^A+gu^A,\Phi^*_A+g v^A]
=S^0[\Phi^A,\Phi^*_A]+g\left(
\frac{\delta S^0}{\delta\Phi^A}u^A
+\frac{\delta S^0}{\delta\Phi^*_A}v_A
\right)
+\cdots$\,,
which can absorb $(X,S^0)$
by choosing $u^A$ and $v_A$ appropriately.

In this paper, we will choose a cubic interaction
as $S^1$.
In this case, $S^1$
can be expanded in the antighost number as
\begin{align}
S^1=a_2+a_1+a_0
\label{eqn:S1}
\end{align}
where $agh(a_i)=i$.
Let us explain the reason why $a_n~(n>2)$ are excluded.
Since  the action $S$ has ghost number 0, $gh(a_i)=0$, 
$pgh(a_i)=agh(a_i)=i$.
As $pgh(a_3)=3$, $a_3$ must be composed of three $c$s.
But such an $a_3$ has $gh(a_3)\neq 0$.
This is because we exclude $a_3$ in the expansion.
For $a_n~(n>3)$, $pgh(a_n)=n$.
It is impossible to
construct such a term as a cubic interaction.

Similarly, we expand $\delta_B$ 
with respect to $agh$
as
\begin{align}
\delta_B=\Delta+\Gamma~,
\end{align}
where $agh(\Delta)=-1$ and $agh(\Gamma)=0$. 
The $\Delta $ and $\Gamma$
are nilpotent and anticommute each other.
These differentials
act on fields and antifields as summarized in Table.\ref{table:Delta, Gamma}.
\begin{table}[h]
 \centering
 \begin{align*}
\begin{array}{|c|c|c|}\hline 
Z & \Delta(Z) & \Gamma(Z) \\\hline 
\phi_{\mu_1\cdots\mu_s} & 0 & \pa_{(\mu_1}c_{\mu_2\cdots\mu_s)} \\\hline 
c_{\mu_1\cdots\mu_{s-1}} & 0 & 0  \\\hline 
\phi^*_{\mu_1\cdots\mu_s} & -G_{\mu_1\cdots\mu_s}(\phi)& 0  \\\hline 
c^*_{\mu_1\cdots\mu_{s-1}} &{ \displaystyle -s\pa\cdot\phi^*_{\mu_1\cdots\mu_{s-1}} }
{ +\frac{s}{D+2s-6}\eta_{(\mu_1\mu_2}\pa\cdot\phi^*{}'_{\mu_3\cdots\mu_{s-1})}  } & 0  \\\hline 
\end{array}
\end{align*}
   \caption{Action of $\Delta$ and $\Gamma$}
  \label{table:Delta, Gamma}
\end{table}
It is easy to see that \bref{BRST closed} is expanded with respect to $agh$ as
\begin{align}
\Gamma 
a_2&=0~,
\label{eqn: agh 2}\\
\Delta a_2+\Gamma a_1
&=0~,
\label{eqn: agh 1}\\
\Delta a_1+\Gamma a_0
&=0~,
\label{eqn: agh 0}
\\
\Delta a_0&=0~.
\label{eqn: agh -1}
\end{align}
Since $agh(a_0)=0$, $a_0$
does not contain any antifields $\Phi^*_A$.
This implies \bref{eqn: agh -1}.

First, we consider \bref{eqn: agh 2}.
Since $a_2$ has $agh=pgh=2$, it is composed of two ghosts $c$s 
and one anti-ghost $c^*$.
For a non-trivial $a_1$,
$\Delta a_2$
has to be $\Gamma$ exact
as seen from \bref{eqn: agh 1}.
This implies that
$\pa c$ must be included in $a_2$,
such as
$\disp \intd 
c^*_{\mu_1\cdots \mu_p\rho_1\cdots\rho_r}
c_{\nu_1\cdots\nu_q}{}^{\rho_1\cdots\rho_r}
\pa^{(\mu_1}c^{\mu_2\cdots\mu_p\nu_1\cdots\nu_q)}$.
This means that  $a_2$ is $\Gamma$ exact.
As is explained in appendix A,
such a $\Gamma$-exact $a_2$ leads to a BRST-trivial $S^1$.
So, we may set
\begin{align}
a_2=0\,.
\label{eqn:a2}
\end{align}

We construct $a_1$ satisfying $\Gamma a_1=0$.
Since $a_1$ has $agh=pgh=1$,
it is composed of a ghost $c$, an antifield $\phi^*$, and a gauge field $\phi$.
For $\Gamma a_1=0$, $\phi$ should appear as 
a gauge-invariant form
such as $G(\phi),$\footnote{
One may use $F$ instead of $G$,
because the difference between $a_1$ with $G$ and $a_1$ with $F$
is absorbed by a field-redefinition.
In fact, the difference results in a term
proportional to $\pa\cdot c$
which is just a $\Gamma$-exact term.
}
since $\Gamma G(\phi)=0$ which follows from $\delta_B F=0$.
In addition, 
noting that $\pa_{(\mu_1}c_{\mu_2\cdots\mu_{p}\nu_1\cdots \nu_q)}=\Gamma \phi$ and
$\pa\cdot c_{\mu_2\cdots\mu_{p}\nu_1\cdots \nu_q}=\frac{1}{2} \Gamma \phi'_{\mu_2\cdots\mu_{p}\nu_1\cdots \nu_q}$,
we choose
\begin{align}
&
\intd 
G^{\rho_1\cdots\rho_r\mu_1\cdots\mu_p}(\phi)
\pa_{(\mu_1}c_{\mu_2\cdots\mu_{p})\nu_1\cdots \nu_q}
\phis^{\nu_1\cdots\nu_q}{}_{\rho_1\cdots\rho_r}~
\label{A1}
\end{align}
as $a_1$,
which is not $\Gamma$ exact.
This is the $a_1$ from which all interaction terms are shown to be constructed successively.
In this paper, we are concerned with the special case with $r=0$.
Even for the case with $r\neq0$,
we have found that the $S^n$ can be constructed successively.
We hope that we will report this result in another place \cite{SS2}.

\medskip
In the followings we will construct a BRST-invariant action $S$
containing all orders in $g$.
In the next section,
we consider two interacting gauge fields
with spins $s$ and $2s$.
In section \ref{sec:three fields}, we generalize further  to the case
with three interacting gauge fields
with spins $s_1$, $s_2$, and $(s_1+s_2)$.


\section{
Interaction of two gauge fields with spins $s$ and $2s$
}\label{sec:two fields}

In this section,
we introduce two gauge fields with spins $s$ and $2s$.
The free action for the spin-$s$ gauge field
is given in \bref{eqn:S^0}.
For notational simplicity,
rank-$s$ fields are denoted
as column vectors,
such as
$\phi_{\mu_1\cdots\mu_s}=\phib$,
$\phi^*_{\mu_1\cdots\mu_s}=\phib^*$,
and
$\pa_{(\mu_1} c_{\mu_2\cdots\mu_s)}=\pacb$.
The number of components of these vectors is $d(s)\equiv \frac{(D-1+s)!}{(D-1)!s!}$.
In this notation, the free action is written as
\begin{align}
S^0(s)&=\intd
\left(
\frac{1}{2}\phib^TG(\phib)+
\phib^*{}^T\pacb
\right)\,.
\label{eqn:S^0 s}
\end{align}
The rank-$2s$ fields are represented as $d(s)\times d(s)$ square matrices,
such as
$\phi_{\mu_1\cdots\mu_{2s}}=\phi$,
$\phi^*_{\mu_1\cdots\mu_{2s}}=\phi^*$
and
$\pa_{(\mu_1} c_{\mu_2\cdots\mu_{2s})}=\pa c$,
so that the free action is expressed as
\begin{align}
S^0(2s)&=\intd\tr\left( 
\frac{1}{2}\phi G(\phi)
+\phi^*{}\pa c
\right)\,.
\label{eqn:S^0 2s}
\end{align}
The total free action is
a sum of \bref{eqn:S^0 s} and  \bref{eqn:S^0 2s}.

\subsection{Three-point interaction}\label{sec:S1}

As explained in section \ref{sec:BRST deformation},
we choose the $\Gamma$-nontrivial $a_1$ as
\begin{align}
a_1=\intd G(\phib)^T\tilde \pa c \phib^*~,
\label{eqn:a1}
\end{align}
which is \bref{A1} with $r=0$.
We will show that
all higher-order terms can be constructed from this $a_1$.
Here we have introduced $\tilde \pa c=\pa_{(\mu_1}c_{\mu_2\cdots\mu_s)\nu_1\cdots \nu_s}$.
We note that $\pa c$ in \bref{eqn:S^0 2s} is a symmetric matrix
but $\tilde\pa c$ is not.
In fact, we note that $G(\phib)^T\pa c \phib^*=G(\phib)^T(\tilde\pa c+(\tilde\pa c)^T) \phib^*
\neq G(\phib)^T\tilde \pa c \phib^*$.
One may choose as $a_1$
\begin{align}
\intd \phib^* {}^T\tilde \pa c G(\phib)
=-
\intd G(\phib)^T(\tilde \pa c)^T \phib^*
\label{eqn:a1 instead}
\end{align}
instead of \bref{eqn:a1},
so that
the obtained action $S$ should be the same
subject to a field redefinition.
This is because
the difference  between 
\bref{eqn:a1} and \bref{eqn:a1 instead}
is a $\Gamma$-exact term,
$\intd G(\phib)^T\pa c \phib^*=-\Gamma\intd G(\phib)^T\phi \phib^*$.

\medskip
It is obvious that \bref{eqn:a1} 
satisfies $\Gamma a_1=0$
as $\Gamma G(\phib)=0$
which follows from the gauge invariance of $F(\phib)$.
To solve \bref{eqn: agh 0} for $a_0$,
we derive
\begin{align}
-\Delta a_1=\intd G(\phib)^T\tilde \pa c G(\phib)
=\intd \frac{1}{2}G(\phib)^T\pa c G(\phib)
=\Gamma \intd \frac{1}{2}G(\phib)^T\phi G(\phib)\,.
\end{align}
This  implies that
we may determine 
$a_0$
as
\begin{align}
a_0=\intd \frac{1}{2}G(\phib)^T\phi G(\phib)\,.
\label{eqn:a0}
\end{align}
As a result,
we have obtained $S^1$ in \bref{eqn:S1}
composed of \bref{eqn:a2},  \bref{eqn:a1}, and \bref{eqn:a0}.

\subsection{Four-point interaction}\label{sec:S2}
We proceed to derive $S^2$,
which is determined by the master equation \bref{eqn:master equation g^n}
for $n=2$,
\begin{align}
2(S^2,S^0)+(S^1,S^1)=0
~~\leftrightarrow~~
\delta_BS^2+\frac{1}{2}(S^1,S^1)=0~.
\label{eqn:master equation n=2}
\end{align}
This implies that $S^2$ is composed of four fields because $S^1$ is composed of three fields.
Expanding $S^2$ with respect to $agh$ as
\begin{align}
S^2=b_2+b_1+b_0\,
\label{eqn:S2}
\end{align}
where $agh(b_i)=i$,
we find that \bref{eqn:master equation n=2}
reduces to
\begin{align}
\Gamma b_2=&0\,,
\label{eqn:ME S2 agh 2}\\
\Delta b_2 +\Gamma b_1
+\frac{1}{2}(a_1,a_1)=&0\,,
\label{eqn:ME S2 agh 1}\\
\Delta b_1 +\Gamma b_0
+(a_1,a_0)=&0\,.
\label{eqn:ME S2 agh 0}
\end{align}
We have not included $b_n$ ($n>2$) in the expansion in \bref{eqn:S2}
simply because
we can solve \bref{eqn:master equation n=2}
without them.

First, we will solve \bref{eqn:ME S2 agh 1}
for $b_2$ and $b_1$.
One derives,
using $a_1$ in \bref{eqn:a1},
\begin{align}
-\frac{1}{2}(a_1,a_1)
=&\intd G(\phib)^T\tilde\pa c \Gt(\tilde\pa c\phib^*)
\nn\\=&
\intd G(\phib)^T\pa c \Gt(\tilde\pa c\phib^*)
-\intd G(\phib)^T(\tilde\pa c)^T \Gt(\tilde\pa c\phib^*)~.
\end{align}
In the second equality, we have used $\pa c=\pat c+(\pat c)^T$.
We find that 
the first term in the last line is $\Gamma$-exact
\begin{align}
\Gamma \intd G(\phib)^T\phi \Gt(\tilde\pa c\phib^*)\,,
\end{align}
while the second term is $\Delta$ exact
\begin{align}
-\intd (\tilde\pa c G(\phib))^T \Gt(\tilde\pa c\phib^*)
=\Delta \intd \frac{1}{2} (\tilde\pa c \phib^*)^T \Gt( \tilde\pa c\phib^*)~,
\end{align}
where we have used \bref{eqn:Gt}.
As a result, we obtain
\begin{align}
b_2=& \intd \frac{1}{2} (\tilde\pa c \phib^*)^T \Gt( \tilde\pa c\phib^*)\,,
\label{eqn:b2}\\
b_1=&\intd G(\phib)^T\phi \Gt(\tilde\pa c\phib^*)\,.
\label{eqn:b1}
\end{align}
It is obvious to see that $b_2$ above satisfies \bref{eqn:ME S2 agh 2}.

Next, we will solve \bref{eqn:ME S2 agh 0} for $b_0$.
Noting that
\begin{align}
-(a_1,a_0)=&\intd G(\phib)^T\tilde\pa c \Gt(\phi G(\phib))
\nn\\=&
\intd  \Gt(\phi G(\phib))^T (\tilde\pa c)^TG(\phib)
\nn\\
=&\intd (\phi G(\phib))^T\Gt((\tilde\pa c)^TG(\phib))\,,
\end{align}
where \bref{eqn:Gt} is used in the last equality,
we find that
\begin{align}
-\Delta b_1-(a_1,a_0)
=&\intd G(\phib)^T\phi\Gt(\tilde\pa cG(\phib))
+\intd (\phi G(\phib))^T\Gt((\tilde\pa c)^TG(\phib))
\nn\\=&
\intd G(\phib)^T\phi\Gt(\pa cG(\phib))
\nn\\=&
\Gamma \intd \frac{1}{2}(\phi G(\phib))^T\Gt(\phi G(\phib))~.
\end{align}
This implies that
\begin{align}
b_0=\intd \frac{1}{2}(\phi G(\phib))^T\Gt(\phi G(\phib))~.
\label{eqn:b0}
\end{align}
As a result, we found that 
$S^2$ in \bref{eqn:S2} is  composed of
\bref{eqn:b2}, \bref{eqn:b1}, and \bref{eqn:b0}.

\subsection{Five-point interaction}\label{sec:S3}

In order to guess the form of $S^n$,
let us proceed to derive $S^3$,
which is determined by the master equation \bref{eqn:master equation g^n},
for $n=3$
\begin{align}
(S^3,S^0)+(S^2,S^1)=0
~~\leftrightarrow~~
\delta_BS^3+(S^1,S^2)=0~.
\label{eqn:master equation n=3}
\end{align}
Expanding $S^3$ with respect to $agh$ as
\begin{align}
S^3=c_2+c_1+c_0\,
\label{eqn:S3}
\end{align}
where $agh(c_i)=i$,
we find that \bref{eqn:master equation n=3}
reduces to
\begin{align}
\Gamma c_2+(a_1,b_2)=&0\,,
\label{eqn:ME S3 agh 2}\\
\Delta c_2 +\Gamma c_1
+(a_1,b_1)
+(a_0,b_2)=&0\,,
\label{eqn:ME S3 agh 1}\\
\Delta c_1 +\Gamma c_0
+(a_1,b_0)+(a_0,b_1)=&0\,.
\label{eqn:ME S3 agh 0}
\end{align}
As seen below,
we can solve \bref{eqn:master equation n=3}
without $c_n$ ($n>2$),
so that 
they are not included in \bref{eqn:S3}.
We first solve \bref{eqn:ME S3 agh 2} for $c_2$.
One derives,
using  \bref{eqn:a1} and \bref{eqn:b2},
\begin{align}
-(a_1,b_2)
=&\intd \Gt(\pat c \phib^*)^T \pat c \Gt(\pat c \phib^*)
=
\intd\frac{1}{2} \Gt(\pat c\phib^*)^T \pa c \Gt(\pat c \phib^*)
\nn\\=&
\Gamma \intd\frac{1}{2} \Gt(\pat c\phib^*)^T \phi \Gt(\pat c \phib^*)
~
\end{align}
where in the second equality, $\pat c+(\pat c)^T=\pa c$ is used.
This implies that
\begin{align}
c_2= \intd\frac{1}{2} \Gt(\pat c\phib^*)^T \phi \Gt(\pat c \phib^*)\,.
\label{eqn:c2}
\end{align}
Next, we solve \bref{eqn:ME S3 agh 1}
for $c_1$.
For this purpose we derive
\begin{align}
-\Delta c_2=&
\intd \Gt(\pat c \phib^*)^T \phi\Gt(\pat cG(\phib))\,,
\label{eqn:Delta c2}\\
-(a_1, b_1)=&
\intd [ G(\phib)^T\pat c \Gt(\phi\Gt(\pat c\phib^*))
+\Gt(\phi G(\phib))^T \pat c \Gt(\pat c\phib^*)
]\,,
\label{eqn:(a1,b1)}\\
-(a_0,b_2)=&
\intd \Gt(\pat c\phib^*)^T \pat c \Gt(\phi G(\phib))\,.
\label{eqn:(a0,b2)}
\end{align}
Combining the second term on the right-hand side of \bref{eqn:(a1,b1)}
with the right-hand side of \bref{eqn:(a0,b2)},
we obtain
\begin{align}
&
\intd \Gt(\phi G(\phib))^T\pa c \Gt(\pat c \phib^*)\nn\\&
=\Gamma\intd \Gt(\phi G(\phib))^T\phi \Gt(\pat c \phib^*)
-\intd \Gt(\pa c G(\phib))^T\phi \Gt(\pat c \phib^*)\,.
\label{eqn:S3 A}
\end{align}
On the other hand, the first term on the right-hand side of \bref{eqn:(a1,b1)}
may be rewritten as
\begin{align}
\intd \Gt(\phi\Gt(\pat c\phib^*))^T (\pat c )^TG(\phib)
=\intd \Gt(\pat c\phib^*)^T \phi\Gt( (\pat c )^TG(\phib))\,,
\end{align}
so that this term and the right-hand side of \bref{eqn:Delta c2}
make
\begin{align}
\intd \Gt(\pat c\phib^*)^T \phi\Gt(\pa c G(\phib))\,.
\label{eqn:S3 B}
\end{align}
Because \bref{eqn:S3 B} cancels out the last term in \bref{eqn:S3 A},
we conclude that
\begin{align}
-\Delta c_2-(a_1,b_1)-(a_0,b_2)
=\Gamma\intd \Gt(\phi G(\phib))^T\phi \Gt(\pat c \phib^*)\,,
\end{align}
which implies that
\begin{align}
c_1=\intd \Gt(\phi G(\phib))^T\phi \Gt(\pat c \phib^*)\,.
\label{eqn:c1}
\end{align}
Finally, \bref{eqn:ME S3 agh 0} is examined to determine $c_0$.
It is straightforward to see that
\begin{align}
-\Delta c_1-(a_1,b_0)=&
\intd \Gt(\phi G(\phib))^T\phi\Gt(\pat c G(\phib))
+\intd G(\phib)^T\pat c \Gt(\phi \Gt(\phi G(\phib)))
\nn\\
=&\intd \Gt (\phi G(\phib))^T \phi \Gt(\pa c G(\phib))
\nn\\
=&\Gamma \intd \frac{1}{2} \Gt (\phi G(\phib))^T \phi \Gt(\phi G(\phib))
-\intd \frac{1}{2} \Gt (\phi G(\phib))^T \pa c \Gt(\phi G(\phib))
\label{eqn:S3 C}
\end{align}
where in the second equality we have used \bref{eqn:Gt} and transposed the integrand,
and that
\begin{align}
-(a_0,b_1)=&\intd \Gt(\phi G(\phib))^T \pat c \Gt(\phi G(\phib))
\nn\\
=&\intd \frac{1}{2} \Gt(\phi G(\phib))^T \pa c \Gt(\phi G(\phib))\,.
\label{eqn:S3 D}
\end{align}
Because \bref{eqn:S3 D}
cancels out the last term in the right-hand side of \bref{eqn:S3 C},
we conclude that
\begin{align}
-\Delta c_1 -(a_1,b_0)-(a_0,b_1)=
\Gamma \intd \frac{1}{2} \Gt (\phi G(\phib))^T \phi \Gt(\phi G(\phib))\,.
\end{align}
This implies that
\begin{align}
c_0= \intd \frac{1}{2} \Gt (\phi G(\phib))^T \phi \Gt(\phi G(\phib))\,.
\label{eqn:c0}
\end{align}
As a result,
$S^3$ in \bref{eqn:S3}
is found to be composed of
\bref{eqn:c2}, \bref{eqn:c1}, and \bref{eqn:c0}.

\subsection{$n$-point interaction}

From the results obtained in the subsections \ref{sec:S1}, \ref{sec:S2}, and \ref{sec:S3},
we can guess the form of $S^n$ as follows
\begin{align}
S^n=&\alpha_2^n+\alpha_1^n+\alpha_0^n\,,
\label{eqn:Sn}\\
\alpha_2^n=&\intd \frac{1}{2}(\pat c\phib^*)^T\Phi^{n-2}[\Gt(\pat c\phib^*)]
=\intd \frac{1}{2}\Phi^{n-2} [\Gt(\pat c \phib^*)]^T\pat c \phib^*\,,
\label{eqn:Sn alpha2}\\
\alpha_1^n=&\intd (\pat c\phib^*)^T\Phi^{n-1}[G(\phib)]
=\intd \phib^T \Phi^{n-1}[\Gt(\pat c\phib^*)]\,,
\label{eqn:Sn alpha1}\\
\alpha_0^n=&\intd\frac{1}{2}\phib^T \Phi^{n}[G(\phib)]\,.
\label{eqn:Sn alpha0}
\end{align}
Here we have introduced $\Phi$ by $\Phi^0[A]=A$, 
$\Phi[A]\equiv \Gt(\phi A)$,
$\Phi^2[A]\equiv \Gt(\phi\Gt(\phi A))$, and so on.
Since $\alpha^1_2=a_2=0$ for $S^1$,
$\Phi^{-1}[A]=0$ is understood.
Observe that $S^1, \,S^2$, and $S^3$
derived in the previous subsections,
coincide with $S^n$ in \bref{eqn:Sn}-\bref{eqn:Sn alpha0} for $n=1,2,3$,
respectively.

We will show that
$S^n$ ($n\geq1$) above solves
the master equation \bref{eqn:master equation g^n}.
Now, suppose that
$S^k$ ($k<n$)
solves the master equation at the order of $g^k$
($k<n$).
We will solve
the master equation \bref{eqn:master equation g^n}
for $\alpha_2^n$, $\alpha_1^n$, and $\alpha_0^n$
and show that they coincide with
\bref{eqn:Sn alpha2}, \bref{eqn:Sn alpha1}, and \bref{eqn:Sn alpha0}.

The master equation \bref{eqn:master equation g^n}
at the order of $g^n$
is expanded with respect to  $agh$
as
\begin{align}
&\Gamma \alpha^n_2+\frac{1}{2}\sum_{k=2}^{n-1}(\alpha_2^k,\alpha_1^{n-k})
+\frac{1}{2}\sum_{k=1}^{n-2}(\alpha_1^k,\alpha_2^{n-k})=0\,,
\label{eqn:ME Sn 2}\\
&
\Delta \alpha^n_2+\Gamma \alpha^n_1
+\frac{1}{2}\sum_{k=2}^{n-1}(\alpha_2^k,\alpha_0^{n-k})
+\frac{1}{2}\sum_{k=1}^{n-1}(\alpha_1^k,\alpha_1^{n-k})
+\frac{1}{2}\sum_{k=1}^{n-2}(\alpha_0^k,\alpha_2^{n-k})
=0\,,
\label{eqn:ME Sn 1}\\
&
\Delta \alpha^n_1+\Gamma \alpha^n_0
+\frac{1}{2}\sum_{k=1}^{n-1}(\alpha_1^k,\alpha_0^{n-k})
+\frac{1}{2}\sum_{k=1}^{n-1}(\alpha_0^k,\alpha_1^{n-k})
=0\,.
\label{eqn:ME Sn 0}
\end{align}
First we will solve \bref{eqn:ME Sn 2}
for $\alpha_2^n$.
Since
\begin{align}
(\alpha_2^k,\alpha_1^{n-k})=&
-\intd \Phi^{k-2}[\Gt(\pat c\phib^*)]^T\pat c\Phi^{n-k-1}[\Gt(\pat c\phib^*)]\,,
\\
(\alpha_1^k,\alpha_2^{n-k})=&
-\intd \Phi^{n-k-2}[\Gt(\pat c\phib^*)]^T\pat c\Phi^{k-1}[\Gt(\pat c\phib^*)]\,,
\end{align}
we obtain
\begin{align}
-\frac{1}{2}\sum_{k=2}^{n-1}(\alpha_2^k,\alpha_1^{n-k})
-\frac{1}{2}\sum_{k=1}^{n-2}(\alpha_1^k,\alpha_2^{n-k})
=&\frac{1}{2}\intd \sum_{l=0}^{n-3} \Phi^l[\Gt(\pat c\phib^*)]^T\pa c\Phi^{n-3-l}[\Gt(\pat c\phib^*)]
\nn\\=&
\Gamma \intd   \frac{1}{2}(\pat c\phib^*)^T\Phi^{n-2}[\Gt(\pat c \phib^*)]
\,.
\end{align}
This implies that $\alpha_2^n$ is given as \bref{eqn:Sn alpha2}.

Next, we derive $\alpha_1^n$ from \bref{eqn:ME Sn 1}.
We observe
that
\begin{align}
&-\Delta \alpha_2^n
-\frac{1}{2} \sum_{k=2}^{n-1}(\alpha_2^k,\alpha_0^{n-k})
-\frac{1}{2} \sum_{k=1}^{n-2}(\alpha_0^k,\alpha_2^{n-k})
\nn\\
&=\intd
(\pat c\phib^*)^T\Phi^{n-2}[\Gt(\pat c G)]
+\intd \sum_{l=0}^{n-3}\Phi^l[\Gt(\pat c\phib^*)]^T \pat c\Phi^{n-2-l}[G(\phib)]
\nn\\&=
\intd \sum_{l=0}^{n-2}\Phi^l[\Gt(\pat c\phib^*)]^T \pat c\Phi^{n-2-l}[G(\phib)]\,
\end{align}
where in the second equality
we have used
the useful relation
\begin{align}
\intd A^T\Phi^m[\Gt(B)]=\intd \Phi^m[\Gt(A)]^TB
\end{align}
which follows from \bref{eqn:Gt}.
Furthermore, deriving
\begin{align}
-\frac{1}{2}\sum_{k=1}^{n-1}(\alpha_1^k,\alpha_1^{n-k})
=\intd\sum_{l=0}^{n-2}\Phi^l[G(\phib)]^T\pat c \Phi^{n-2-l}[\Gt(\pat c \phib^*)]\,,
\end{align}
we find that
\begin{align}
&
-\Delta \alpha_2^n
-\frac{1}{2} \sum_{k=2}^{n-1}(\alpha_2^k,\alpha_0^{n-k})
-\frac{1}{2}\sum_{k=1}^{n-1}(\alpha_1^k,\alpha_1^{n-k})
-\frac{1}{2} \sum_{k=1}^{n-2}(\alpha_0^k,\alpha_2^{n-k})
\nn\\&=
\intd \sum_{l=0}^{n-2}\Phi^l[\Gt(\pat c\phib^*)]^T \pa c\Phi^{n-2-l}[G(\phib)]
\nn\\&=
\Gamma \intd (\pat c \phib^*)^T\Phi^{n-1}[G(\phib)]
\,.
\end{align}
This implies that $\alpha_1^n$ is given as in \bref{eqn:Sn alpha1}.

Finally, we derive $\alpha_0^n$ from \bref{eqn:ME Sn 0}.
It is straightforward to derive
\begin{align}
&
-\Delta \alpha^n_1
-\frac{1}{2}\sum_{k=1}^{n-1}(\alpha_1^k,\alpha_0^{n-k})
-\frac{1}{2}\sum_{k=1}^{n-1}(\alpha_0^k,\alpha_1^{n-k})
\nn\\&=
\intd
\Phi^{n-1}[G(\phib)]^T\pat c G(\phib)
+\intd
\sum_{l=0}^{n-2}\Phi^l[G(\phib)]^T\pat c \Phi^{n-k}[G(\phib)]
\nn\\&=
\intd \sum_{l=0}^{n-1}\Phi^l[G(\phib)]^T\pat c \Phi^{n-k}[G(\phib)]
\nn\\&=
\intd \frac{1}{2}\sum_{l=0}^{n-1}\Phi^l[G(\phib)]^T\pa c \Phi^{n-k}[G(\phib)]
\nn\\&=
\Gamma \intd \frac{1}{2}\phib^T\Phi^{n}[G(\phib)]\,.
\end{align}
This implies that 
$\alpha_0^n$ is given as in \bref{eqn:Sn alpha0}.

Summarizing the above results,
we have shown that $S^n$ is definitely given in \bref{eqn:Sn}-\bref{eqn:Sn alpha0}.

\subsection{BRST-invariant action of interacting spin-$s$ and spin-$2s$ gauge fields}

We examine the total action $S$.
In the above, we derived $S^k$ ($k=1,2,\ldots$).
The free action is a sum of $S^0(s)$ and $S^0(2s)$ given in \bref{eqn:S^0 s} and \bref{eqn:S^0 2s},
respectively.
By gathering these results together, the total action is given as
\begin{align}
S=&S^0(s)+S^0(2s)+\sum_{k=1}^\infty g^k S^k
\nn\\=&
\intd \Big[
\frac{1}{2}\phib^TG(\phib)+
\phib^*{}^T\pacb
+\tr( 
\frac{1}{2}\phi G(\phi)
+\phi^*{}\pa c
)
+\frac{1}{2}\phib^T\sum_{k=1}g^k\Phi^k[G(\phib)]
\nn\\&\hspace{13mm}
+\phib^T\sum_{k=1}g^k\Phi^{k-1}[\Gt(\pat c \phib^*)]
+\frac{1}{2}(\pat c\phib^*)^T\sum_{k=2}g^k\Phi^{k-2}[\Gt(\pat c \phib^*)]
\Big]\,.
\end{align}
We find that the action turns into the form
expanded in $agh$
as
\begin{align}
S=&S_0+S_1+S_2\,,\\
S_0=&\intd\Big[
\frac{1}{2} \phib^T\frac{1}{1-g\Phi}G(\phib)
+\tr\left(\frac{1}{2}\phi G(\phi)\right)
\Big]\,,
\label{eqn:action S_0}
\\
S_1=&\intd\Big[
\phib^*{}^T\pacb
+\tr\left(
\phi^*{}\pa c
\right)
+\phib^T \frac{g}{1-g\Phi}\Gt(\pat c \phib^*)
\Big]\,,\\
S_2=&\intd
\frac{1}{2}(\pat c\phib^*)^T\frac{g^2}{1-g\Phi} \Gt(\pat c \phib^*)\,,
\end{align}
where $agh(S_i)=i$.
This is one of the main results of this paper.
The obtained action $S$ contains all orders in $g$,
and BRST-invariant $\delta_B^g S=(S,S)=0$.
It is not likely that $S$ and $S^0$ are related to each other by a field
redefinition as far as we can tell.

\medskip
We have obtained BRST-invariant action of interacting gauge fields
by using the BRST-antifield formalism.
Here we examine the gauge invariance of the action $S_0$.
The gauge transformation can be extracted  from the BRST transformation
$\delta_B^g X=(X,S)$.
We find that
the gauge transformation with a rank-$(s-1)$ parameter $\xib$
remains unchanged,
\begin{align}
\delta \phib=\pa \xib\,,~~~
\delta \phi=0\,,
\label{eqn:gauge tfn1}
\end{align}
while the gauge transformation with  a rank-$(2s-1)$ parameter $\xi$ 
turns into
\begin{align}
\delta \phib=-\left(\left[
\frac{g}{1-g\Phi}G(\phib)
\right]^T\pat \xi
\right)^T
~,~~~
\delta \phi=\pa\xi
\,.
\label{eqn:gauge trans s}
\end{align}
We note that
higher-order interactions make $\delta \phib$ non-trivial.
It is instructive to check the gauge invariance of $S_0$
in \bref{eqn:action S_0}
under \bref{eqn:gauge trans s}.
The gauge variation of $S_0$ is
\begin{align}
\delta S_0=&\intd\left[
\frac{1}{2} \phib^T\delta ( \frac{1}{1-g\Phi} )G(\phib)
-\left(
\frac{g}{1-g\Phi}G(\phib)
\right)^T\pat \xi\frac{1}{1-g\Phi} G(\phib)
\right]\,.
\label{eqn:S_0 variation}
\end{align}
The gauge variation of the  second term on the right-hand side of  \bref{eqn:action S_0}
is eliminated by $G(\pa\xi)=0$.
We find that the first term on the right-hand side of \bref{eqn:S_0 variation}
turns into 
\begin{align}
&\sum_{k=0}^\infty\intd 
\frac{1}{2} \sum_{l=0}^{k-1}(g\Phi)^l[G(\phib)]^T(g\pa \xi)
(g\Phi)^{k-l-1}[G(\phib)]
\nn\\&
=
\intd 
\frac{1}{2} \sum_{l=0}^\infty (g\Phi)^l[G(\phib)]^T(g\pa \xi)
\sum_{k=0}^\infty (g\Phi)^{k}[G(\phib)]
\nn\\&
=
\intd 
\frac{1}{2} \left(\frac{1}{1-g\Phi} G(\phib)\right)^T(g\pa \xi)
\frac{1}{1-g\Phi} G(\phib)
\nn\\&
=
\intd 
\left(\frac{1}{1-g\Phi} G(\phib)\right)^T(g\pat \xi)
\frac{1}{1-g\Phi} G(\phib)~.
\end{align}
It follows that the right-hand side of \bref{eqn:S_0 variation}
cancels out.
Summarizing, we find that the action $S_0$ in \bref{eqn:action S_0}
is invariant under the gauge transformations in \bref{eqn:gauge tfn1} and \bref{eqn:gauge trans s}.
\medskip

In section \ref{sec:AdS}, we show that
the action $S$ obtained in this section
can be easily generalized to the action on  AdS spaces.

\section{
Interaction of three gauge fields with spins $s_1$, $s_2$, and $(s_1+s_2)$
}\label{sec:three fields}

In this section we will slightly generalize the model examined in the previous section.
We introduce three gauge fields
with spins $s_1$, $s_2$, and $(s_1+s_2)$.
The spin-$s_I$ ($I=1,2$) fields are denoted  as $d(s_I)$-component column vectors,
such as
$\phi_{\mu_1\cdots \mu_{s_I}}=\phib^{(I)}$,
$\phis_{\mu_1\cdots \mu_{s_I}}=\phib^*{}^{(I)}$
and $\pa_{(\mu_1}c_{\mu_2\cdots\mu_{s_I})}=\pacb^{(I)}$,
while the spin-($s_1+s_2$) fields are
denoted as rectangular matrices,
$\phi_{\mu_1\cdots \mu_{s_1+s_2}}=\phi$, $\phis_{\mu_1\cdots \mu_{s_1+s_2}}=\phi^*$ and 
$\pa_{(\mu_1}c_{\mu_2\cdots\mu_{s_1+s_2})}=\pa c$.
In this notation,
the free action is written as
\begin{align}
S^0=\intd \Bigg[
\frac{1}{2}
\phib^{(I)}{}^TG(\phib^{(I)})+\phibs^{(I)}{}^T\pacb^{(I)}
+\tr\left( 
\frac{1}{2}\phi G(\phi)
+\phi^*{}\pa c
\right)
\Bigg]\,
\label{eqn:free gene}
\end{align}
where the summation over $I=1,2$ is understood.

As explained above, $S^1$ is expanded as in \bref{eqn:S1} with $a_2=0$.
In the present case, we choose
\begin{align}
a_1=\intd
\Big[&
G(\phib^{(1)})^{\mu_1\cdots\mu_{s_1}}
\pa_{(\mu_1} c_{\mu_2\cdots\mu_{s_1})\nu_1\cdots\nu_{s_2}} 
\phib^*{}^{(2)}{}^{\nu_1\cdots\nu_{s_2}}
\nn\\&
+G(\phib^{(2)})^{\nu_1\cdots\nu_{s_2}}
\pa_{(\nu_1} c_{\nu_2\cdots\nu_{s_2})\mu_1\cdots\mu_{s_1}} 
\phib^*{}^{(1)}{}^{\mu_1\cdots\mu_{s_1}}
\Big]\,.
\label{eqn:a1 gene indices}
\end{align}
We will show that this $a_1$ leads to an action of three interacting gauge fields.
For notational simplicity,
we write \bref{eqn:a1 gene indices} as
\begin{align}
a_1=\intd \left[s_{IJ}
G(\phib^{(I)})^T\pat c \phib^*{}^{(J)}
\right]\,,
\label{eqn:a1 general}
\end{align}
where $s_{IJ}$ denotes a $2\times 2$ matrix of $s_{12}=s_{21}=1$ and $s_{11}=s_{22}=0$.
In this notation,
the index of the derivative in $\pat c$
is always contracted with one of the indices of the field
sitting to its immediate left.
In other words, $\pat c$ in \bref{eqn:a1 general}
is a $d(s_I)\times d(s_J)$ matrix.


We can derive 
$S^1$, $S^2$, and $S^3$
in a similar way to the preceding section.
For this paper to remain self-contained, we give a brief derivation
in appendix \ref{app:derivation}.
From the results obtained there, we can guess the form of $S^n$ ($n\geq 1$) as
\begin{align}
S^n=&\alpha_2^n+\alpha_1^n+\alpha_0^n\,,
\label{eqn:Sn gene}\\
\alpha_2^n=&\intd  \frac{1}{2} s_{IJ}^n \Phi^{n-2}[\Gt(\pat c \phibs^{(I)})]^T\pat c \phibs^{(J)}\,,
\label{eqn:Sn alpha2 gene}\\
\alpha_1^n=&
\intd s_{IJ}^n \phib^{(I)}{}^T \Phi^{n-1}[\Gt(\pat c\phibs^{(J)})]\,,
\label{eqn:Sn alpha1 gene}\\
\alpha_0^n=&\intd\frac{1}{2}s_{IJ}^n\phib^{(I)}{}^T \Phi^{n}[G(\phib^{(J)})]\,,
\label{eqn:Sn alpha0 gene}
\end{align}
where $\Phi^{-1}=0$ is understood because $ \alpha_2^1=a_2=0$.
We note that $s^2_{IJ}=\delta_{IJ}$,
$s^3_{IJ}=s_{IJ}$, and so on.
It is easy to see that $S^n$ for $n=1,2,3$ coincide with those obtained in appendix \ref{app:derivation}.
We will derive $S^n$
supposing that 
$S^k$ ($k<n$)
solve the master equation at the order of $g^k$ ($k<n$).

The master equation \bref{eqn:master equation g^n}
is expanded at the order of $g^n$
as \bref{eqn:ME Sn 2},
\bref{eqn:ME Sn 1},
and 
\bref{eqn:ME Sn 0}.
First, we solve \bref{eqn:ME Sn 2} for $\alpha_2^n$.
It is straightforward to see that
\begin{align}
-\frac{1}{2}\sum_{k=2}^{n-1}(\alpha_2^k,\alpha_1^{n-k})
-\frac{1}{2}\sum_{k=1}^{n-2}(\alpha_1^k,\alpha_2^{n-k})
=&\intd  \frac{1}{2}s_{IJ}^n
\sum_{l=0}^{n-3} \Phi^l[\Gt(\pat c\phibs^{(I)})]^T\pa c\Phi^{n-3-l}[\Gt(\pat c\phibs^{(J)})]
\nn\\=&
\Gamma \intd \frac{1}{2} s_{IJ}^n \Phi^{n-2}[\Gt(\pat c \phibs^{(I)})]^T\pat c \phibs^{(J)}
\,.
\end{align}
This implies that
\begin{align}
\alpha_2^n= \intd \frac{1}{2} s_{IJ}^n \Phi^{n-2}[\Gt(\pat c \phibs^{(I)})]^T\pat c \phibs^{(J)}\,,
\end{align}
which coincides with \bref{eqn:Sn alpha2 gene}.

Next, we  derive $\alpha_1^n$ from \bref{eqn:ME Sn 1}.
For this purpose, we derive
\begin{align}&
-\Delta \alpha_2^n
-\frac{1}{2} \sum_{k=2}^{n-1}(\alpha_2^k,\alpha_0^{n-k})
-\frac{1}{2} \sum_{k=1}^{n-2}(\alpha_0^k,\alpha_2^{n-k})
\nn\\&=\intd s^n_{IJ}\sum_{l=0}^{n-2}
\Phi^l[\Gt(\pat c \phibs^{(I)})]^T\pat c
\Phi^{n-2-l}[G(\phib^{(J)})]\,,
\end{align}
and
\begin{align}
-\frac{1}{2}\sum_{k=1}^{n-1}(\alpha_1^k,\alpha_1^{n-k})
=\intd s^n_{IJ}\sum_{l=0}^{n-2}
\Phi^l[G(\phib^{(I)})]^T\pat c 
\Phi^{n-2-l}[\Gt(\pat c \phibs^{(J)})]\,.
\end{align}
Gathering these results together,
we find
\begin{align}
&
-\Delta \alpha_2^n
-\frac{1}{2} \sum_{k=2}^{n-1}(\alpha_2^k,\alpha_0^{n-k})
-\frac{1}{2}\sum_{k=1}^{n-1}(\alpha_1^k,\alpha_1^{n-k})
-\frac{1}{2} \sum_{k=1}^{n-2}(\alpha_0^k,\alpha_2^{n-k})
\nn\\&=
\intd s^n_{IJ}\sum_{l=0}^{n-2}\Phi^l[\Gt(\pat c\phibs^{(I)})]^T \pa c\Phi^{n-2-l}[G(\phib^{(J)})]
\nn\\&=
\Gamma \intd s^n_{IJ} (\pat c \phibs^{(I)})^T\Phi^{n-1}[G(\phib^{(J)})]
\,.
\end{align}
This implies that $\alpha_1^n$ is given as in \bref{eqn:Sn alpha1 gene}.

Finally, we solve  \bref{eqn:ME Sn 0} for $\alpha_0^n$.
It is now easy
to derive
\begin{align}
&
-\Delta \alpha^n_1
-\frac{1}{2}\sum_{k=1}^{n-1}(\alpha_1^k,\alpha_0^{n-k})
-\frac{1}{2}\sum_{k=1}^{n-1}(\alpha_0^k,\alpha_1^{n-k})
\nn\\&=
\intd s^n_{IJ}\sum_{l=0}^{n-1}\Phi^l[G(\phib^{(I)})]^T\pat c \Phi^{n-k}[G(\phib^{(J)})]
\nn\\&=
\intd \frac{1}{2} s^n_{IJ}\sum_{l=0}^{n-1}\Phi^l[G(\phib^{(I)})]^T\pa c \Phi^{n-k}[G(\phib^{(J)})]
\nn\\&=
\Gamma \intd \frac{1}{2}s^n_{IJ} \phib^{(I)}{}^T\Phi^{n}[G(\phib^{(J)})]\,.
\end{align}
This implies that 
$\alpha_0^n$ is given as in \bref{eqn:Sn alpha0 gene}.

Summarizing the above results,
we have shown that $S^n$ is definitely given as \bref{eqn:Sn gene}-\bref{eqn:Sn alpha0 gene}.

\subsection{BRST-invariant action of interacting
gauge fields
of spins $s_1$, $s_2$, and $s_1+s_2$
}

We examine the total action $S$.
In the above we have derived $S^k$ ($k=1,2,\ldots$).
The free action $S^0$ is given in
\bref{eqn:free gene}.
Gathering these together, we obtain the total action
\begin{align}
S=&S^0+\sum_{k=1}^\infty g^k S^k
\nn\\=&
\intd \Big[
\frac{1}{2}
\phib^{(I)}{}^TG(\phib^{(I)})+\phibs^{(I)}{}^T\pacb^{(I)}
+\tr\left( 
\frac{1}{2}\phi G(\phi)
+\phi^*{}\pa c
\right)
\nn\\&\hspace{13mm}
+\frac{1}{2}\phib^{(I)}{}^T\sum_{k=1}g^ks^k_{IJ}\Phi^k[G(\phib^{(I)})]
+\phib^{(I)}{}^T\sum_{k=1}g^ks^k_{IJ}\Phi^{k-1}[\Gt(\pat c \phibs^{(J)})]
\nn\\&\hspace{13mm}
+\frac{1}{2}(\pat c\phibs^{(I)})^T\sum_{k=2}g^ks^k_{IJ}\Phi^{k-2}[\Gt(\pat c \phibs^{(J)})]
\Big]\,.
\end{align}
It is straightforward to see that the action turns into the form
expanded in $agh$
as
\begin{align}
S=&S_0+S_1+S_2\,\\
S_0=&\intd\Big[
\frac{1}{2} \phib^{(I)}{}^T
\left(\frac{1}{1-gs\Phi}
\right)_{IJ}G(\phib^{(J)})
+\tr\left(\frac{1}{2}\phi G(\phi\right)
\Big]\,,
\label{eqn:action S_0  gene}
\\
S_1=&\intd\Big[
\phibs^{(I)}{}^T\pacb^{(I)}
+\tr(\phi^*{}\pa c)
+\phib^{(I)}{}^T 
\left(\frac{gs}{1-gs\Phi}
\right)_{IJ}\Gt(\pat c \phibs^{(J)})
\Big]\,,\\
S_2=&\intd
\frac{1}{2}\left[
\left(\frac{g^2}{1-gs\Phi}
\right)_{IJ}
\Gt(\pat c \phibs^{(J)})
\right]^T\pat c 
\phibs^{(I)}\,,
\end{align}
where $agh(S_i)=i$.
This action $S$ contains all orders in $g$,
and is invariant under the BRST transformation $\delta_B^g S=(S,S)=0$.
This is one of the main results of this paper.

\medskip
We examine the gauge invariance of the gauge-invariant action $S_0$.
The gauge transformation can be extracted from the BRST transformation
$\delta_B^g X=(X,S)$.
We find that the gauge transformation with a rank-$(s_I-1)$ parameter $\xib^{(I)}$
remains unchanged
\begin{align}
\delta \phib^{(I)}=\pa \xib^{(I)}\,,~~~
\delta \phi=0\,,
\label{eqn:gauge tfn1 gene}
\end{align}
while the gauge transformation with  a rank-$(s_1+s_2-1)$ parameter $\xi$ 
turns into
\begin{align}
\delta \phib^{(I)}
=-\left(\left[\frac{g}{1-gs\Phi}G(\phib)\right]^T\pat \xi\right)^T
~,~~~
\delta \phi=\pa\xi\,.
\label{eqn:gauge trans s gene}
\end{align}
It is instructive to check the gauge invariance of $S_0$
under \bref{eqn:gauge trans s gene}.
The gauge variation of $S_0$ is
\begin{align}
\delta S_0=&\intd\Big[
\frac{1}{2} \phib^{(I)}{}^T\delta \left(\sum_{l=0} (gs_{IJ}\Phi)^l\right) G(\phib^{(J)})
-\left[\frac{1}{1-gs\Phi} G(\phib^{(I)})\right]^T
\pat \xi \frac{gs}{1-gs\Phi} G(\phib^{(J)})
\Big]\,.
\label{eqn:S_0 variation gene}
\end{align}
We find that the first term on the right-hand side of \bref{eqn:S_0 variation gene}
turns into 
\begin{align}
&\sum_{k=0}^\infty\intd 
\frac{1}{2} \sum_{l=0}^{k-1}(gs\Phi)^l[G(\phib^{(I)})]^Tgs\pa \xi
(gs\Phi)^{k-l-1}[G(\phib^{(J)})]
\nn\\&
=
\intd 
\frac{1}{2} \sum_{l=0}^\infty (gs\Phi)^l[G(\phib^{(I)})]^Tgs\pa \xi
\sum_{k=0}^\infty (gs\Phi)^{k}[G(\phib^{(J)})]
\nn\\&
=
\intd 
\left[\frac{1}{1-gs\Phi} G(\phib^{(I)})\right]^T
\pat \xi\frac{gs}{1-gs\Phi} G(\phib^{(J)})\,.
\end{align}
It follows that the right-hand side of \bref{eqn:S_0 variation gene}
cancels out.
As a result, we find that
the action $S_0$ in \bref{eqn:action S_0  gene}
is invariant under the gauge transformations in \bref{eqn:gauge tfn1 gene} and \bref{eqn:gauge trans s gene}.

\medskip

In the next section we write down an action of three interacting gauge fields on AdS spaces
from the action obtained in this section.

\section{Interacting higher-spin gauge fields on AdS spaces}\label{sec:AdS}

We have obtained actions of interacting gauge fields 
in sections \ref{sec:two fields} and \ref{sec:three fields}. 
One advantage of our approach is that
it is easy to generalize the action on a flat spacetime to that on AdS spaces.
In this section, we write down actions of interacting gauge fields on AdS spaces
from those obtained in sections \ref{sec:two fields} and \ref{sec:three fields}.

First of all, we will introduce objects on AdS spaces.
The Fronsdal tensor on AdS spaces is given as
\begin{align}
F_{\mu_1\cdots\mu_s}(\phi)=\,&
\square\phi_{\mu_1\cdots \mu_s}
-\nabla_{(\mu_1}\nabla^\sigma \phi_{\mu_2\cdots\mu_s)\sigma}
+\frac{1}{2}\nabla_{(\mu_1}\nabla_{\mu_2}\phi'_{\mu_3\cdots\mu_s)}
-\frac{m^2}{l^2}\phi_{\mu_1\cdots\mu_s}
-\frac{2}{l^2}g_{(\mu_1\mu_2}\phi'_{\mu_3\cdots\mu_s)}
~,
\label{eqn:Fronsdal AdS}
\end{align}
where $\square=g^{\mu\nu}\nabla_\mu\nabla_\nu$
and $l$ denotes the radius of the AdS space.
The Fronsdal tensor in \bref{eqn:Fronsdal AdS}
 reduces to \bref{Fronsdal flat} in the flat limit
 $l\to \infty$.\footnote{
Note that
 the third term on the right-hand side
contains a factor $1/2$.
This is needed for \bref{eqn:Fronsdal AdS} to reproduce 
\bref{Fronsdal flat}
in the flat limit.
In fact,
the term 
$\frac{1}{2}(\nabla_{\mu_1}\nabla_{\mu_2}+\nabla_{\mu_2}\nabla_{\mu_1})\phi'$
contained on the right-hand side of \bref{eqn:Fronsdal AdS}
becomes $\pa_{\mu_1}\pa_{\mu_2}\phi'$ in the flat limit.
On the other hand,
the right hand side of \bref{Fronsdal flat} contains $\pa_{\mu_1}\pa_{\mu_2}\phi'$,
which coincides with the flat limit of \bref{eqn:Fronsdal AdS}.
}
The $m^2$ means
\begin{align}
m^2=s^2+s(D-6)-2(D-3)~.
\end{align}
Under the 
gauge transformation 
\begin{align}
\delta \phi_{\mu_1\cdots\mu_s}
=\nabla_{(\mu_1}\xi_{\mu_2\cdots\mu_s)}~,
\end{align}
we obtain
\begin{align}
\delta F_{\mu_1\cdots \mu_s}(\phi)=
\frac{1}{2}\nabla_{(\mu_1}\nabla_{\mu_2}\nabla_{\mu_3}\xi'_{\mu_4\cdots\mu_s)}
-\frac{4}{\l^2}g_{(\mu_1\mu_2}\nabla_{\mu_3}\xi'_{\mu_4\cdots \mu_s)}\,.
\end{align}
This reduces to \bref{eqn:delta F flat} in the flat limit.
It follows that the Fronsdal tensor is gauge invariant if $\xi'=0$.
To derive this relation,
we have used the fact that, on AdS spaces,
\begin{align}
[\nabla_{\mu},\nabla_\nu] \phi_{\alpha_1\cdots\alpha_s}
=R_{\mu\nu(\alpha_1}{}^\lambda\phi_{\alpha_2\cdots\alpha_s)\lambda}
~,~~~
R_{\mu\nu\alpha\beta}=-\frac{1}{l^2}(g_{\mu\alpha}g_{\nu\beta}-g_{\nu\alpha}g_{\mu\beta})
~.
\end{align}

A useful and important relation
 is 
\begin{align}
&\int\d^D x\,\sqrt{-g}
\varphi^{\mu_1\cdots\mu_s}\Gt_{\mu_1\cdots\mu_s}(\psi)
=\int\d^D x\,\sqrt{-g}
\Gt_{\mu_1\cdots\mu_s}(\varphi) 
\psi^{\mu_1\cdots\mu_s}\,,
\end{align}
where
\begin{align}
\Gt_{\mu_1\cdots\mu_s}(A)=&
G_{\mu_1\cdots\mu_s}(A) 
+\frac{1}{4}g_{(\mu_1\mu_2}\nabla_{\mu_3}\nabla_{\mu_4}A''_{\mu_5\cdots\mu_s)}
\label{eqn:Gt AdS}
\,,\\
 G_{\mu_1\cdots\mu_s}(A)=&F_{\mu_1\cdots\mu_s}(A)
 -\frac{1}{2}g_{(\mu_1\mu_2} F'_{\mu_3\cdots\mu_s)}(A)
\,. \end{align}
Observe that \bref{eqn:Gt AdS} reduces to \bref{eqn:Gt}
in the flat limit.

Now we will write down the action of interacting gauge fields on AdS spaces.
It is straightforward to obtain the action on AdS spaces from
the action
 obtained in section \ref{sec:two fields}.
 One obtains
\begin{align}
S=&S_0+S_1+S_2\,\\
S_0=&\intd \sqrt{-g}\Big[
\frac{1}{2} \phib^T\frac{1}{1-g\Phi}G(\phib)
+\tr\left(\frac{1}{2}\phi G(\phi)\right)
\Big]\,,
\label{eqn:action S_0 gene AdS}\\
S_1=&\intd\sqrt{-g}\Big[
\phib^*{}^T
\nabla\bm{ c} 
+\tr\left(\phi^*{}
\nabla c \right)
+\phib^T \frac{g}{1-g\Phi}\Gt(
\tilde\nabla c \phib^*)
\Big]\,,\\
S_2=&\intd\sqrt{-g}
\frac{1}{2}(
\tilde \nabla
 c\phib^*)^T\frac{g^2}{1-g\Phi} \Gt(
 \tilde \nabla
  c \phib^*)\,.
\end{align}
On the other hand, from the
action obtained in section \ref{sec:three fields}
one obtains
\begin{align}
S=&S_0+S_1+S_2\,\\
S_0=&\intd\sqrt{-g}\Big[
\frac{1}{2} \phib^{(I)}{}^T
\left(\frac{1}{1-gs\Phi}
\right)_{IJ}G(\phib^{(J)})
+\tr\left(\frac{1}{2}\phi G(\phi)\right)
\Big]\,,
\label{eqn:action S_0 AdS}\\
S_1=&\intd\sqrt{-g}\Big[
\phibs^{(I)}{}^T
\nabla\bm{ c}^{(I)}
+\tr\left(\phi^*{}
\nabla
 c\right)
+\phib^{(I)}{}^T 
\left(\frac{gs}{1-gs\Phi}
\right)_{IJ}\Gt(
\tilde \nabla
 c \phibs^{(J)})
\Big]\,,\\
S_2=&\intd\sqrt{-g}
\frac{1}{2}\left[
\left(\frac{g^2}{1-gs\Phi}
\right)_{IJ}
\Gt(
\tilde \nabla
 c \phibs^{(J)})
\right]^T
\tilde \nabla
 c 
\phibs^{(I)}\,.
\end{align}
These are the actions of interacting gauge fields on AdS spaces.
They contain all orders in $g$, and are  invariant under the BRST transformation
$\delta_B^g S=(S,S)=0$.

\section{Summary and Discussion}

We have constructed actions of interacting bosonic higher-spin gauge fields
in the BRST-antifield formalism.
In section \ref{sec:two fields}, two gauge fields of spins $2s$ and $s$
were examined,
and
three gauge fields of spins $s_1$, $s_2$, and $(s_1+s_2)$
were examined in section \ref{sec:three fields}.
In both cases,
we constructed an action
$S$ as a deformation of the free action $S^0$,
$S=S^0+gS^1+g^2S^2+\cdots$.
Choosing a cubic interaction as $S^1$,
we solved the master equation $(S,S)=0$
from the lower orders in $g$.
We note that our obtained action $S$ contains all orders in $g$
and is BRST invariant, $\delta_B^g S=(S,S)=0$,
while
the free action $S^0$ is BRST invariant, $\delta_B  S^0=(S^0,S^0)=0$.
Our action is composed of terms with $agh\leq 2$,
and may be rearranged as $S=S_0+S_1+S_2$ where $agh(S_i)=i$.
From the BRST transformation $\delta_B^g X=(X,S)$, 
we have extracted the gauge transformation
and
showed the gauge invariance of $S_0$.
In section \ref{sec:AdS},
we wrote down actions of interacting bosonic higher-spin gauge fields on AdS spaces 
from those obtained in sections \ref{sec:two fields} and \ref{sec:three fields}.

\medskip

The obtained action is composed of
$n+2$ fields and $2n+2$ derivatives
at the order of $g^n$.
The cubic interaction term contains four derivatives.
This is compared with models composed of
the
generalized curvature tensor.
These models avoid the no-go theorem
by restricting the number of derivatives.
Our models suffer from the no-go theorem as well.
So the spin of the fields should be restricted appropriately \cite{Metsaev lc cubic}
as follows: 
$s=1,2$ for the $(s,2s,s)$ cubic coupling,
and
$(s_1,s_2,s_3)=(1,1,2), (1,2,3)$, and $(2,2,4)$
for the $(s_1,s_2,s_1+s_2)$ cubic coupling.

Higher-spin gauge theories have attracted renewd interest
in the study of the AdS/CFT correspondence \cite{AdS/CFT}.
Higher-spin gauge theories have been conjectured to be dual to simple conformal field theories.\footnote{
A bosonic higher-spin gauge theory on a three-dimensional AdS space (AdS$_3$)
\cite{3dVasiliev}
has been conjectured \cite{3dV/W} to be dual to  
the 't Hooft limit of the $\CW_N$ minimal model,
while Vasiliev's higher-spin gauge theory on 
AdS$_4$
 \cite{4dVasiliev}
is conjectured \cite{4dV/O} to be dual to the three-dimensional $O(N)$ sigma-model.
}
Our models on AdS spaces are much simpler than them,
and actions are given in this paper.
It will be interesting to explore the CFT duals to 
our models.

In this paper, we have examined a special kind of interaction term
in which each interaction term in $S$
forms an open chain of fields. 
This enabled us to employ a matrix notation for higher-spin fields.
It is possible to use this notation
even if each interaction term forms a closed chain of fields.
In fact, we have found that even in this case we can construct an action of
interacting bosonic higher-spin gauge theory.
We hope to report this result in another place
\cite{SS2}.
The most general interaction term may not form a chain of fields
but a complete graph (or a simplex),
for example a complete graph with four vertices $K_4$ (or a tetrahedron)
for  a four-point interaction.
Each vertex represents a field, and an edge represents a contraction
between two fields.
In this case, we must employ a tensor-like notation.
This is left for future investigation.

Another interesting issue to pursure is to include fermionic
and bosonic gauge fields with mixed indices.
There are some interesting works examining a deformation of the free action with fermions.
In \cite{HLGR fermion BRST-antifield}\cite{Henneaux:2013gba} a systematic analysis including fermionic gauge fields
was performed in the BRST-antifield formalism.\footnote{
The results were shown to be consistent with 
those obtained in the light-cone formulation \cite{Metsaev lc cubic 2}
and with one obtained in the tensionless  limit of string theory
\cite{Sagnotti tensionless limit}.
}
It will be interesting to examine actions of interacting bosonic and fermionic fields
including higher orders in $g$,
and to generalize them to those on AdS spaces.

\section*{Acknowledgments}

The authors would like to thank Takanori Fujiwara, Yoshifumi Hyakutake,
 and Sota Hanazawa
for useful comments. 
They also thank
Taichiro Kugo
for wonderful
and stimulating lectures on the quantum theory of gauge fields given at Ibaraki University.
They also thank the Yukawa Institute for Theoretical Physics at Kyoto University.
Discussions during the YITP Workshop
``Strings and Fields 2019,''
19-23 August  2019
were useful in completing this work.
They would like to acknowledge the organizers
of ``The 29th Workshop on General Relativity and Gravitation in Japan''
held at Kobe University, 25-29 November 2019,
and
the conference
``KEK Theory Workshop 2019" held at the KEK theory center, Tsukuba, 3-6 December 2019,
for their kind hospitality. 
In these workshops, H.S. reported preliminary results summarized in this paper.

\appendix
\section*{Appendices}

\section{$\Gamma$-trivial $a_2$ results in BRST-trivial $S^1$}
\label{app:trivial}

When $S^1$ is BRST trivial,
$S^1=\delta_B m$,
$a_i$ may be expressed as
\begin{align}
a_2&=\Gamma m_2
~,
\label{eqn: trivial agh 2}\\
a_1&=\Delta m_2+\Gamma m_1
~,
\label{eqn: trivial agh 1}\\
a_0&=\Delta m_1+\Gamma m_0
~.
\label{eqn: trivial agh 0}
\end{align}
The reason $\Delta m_3$  is not included in \bref{eqn: trivial agh 2}
is as follows.
As $\Delta m_3$ has $agh=2$ and $gh=0$,
$\Delta m_3$
is composed of a $c^*$ and two $c$s,
but it is not possible to construct $m_3$ from such a $\Delta m_3$.

Suppose that 
$a_2$ is $\Gamma$ exact, as given in 
\bref{eqn: trivial agh 2}.
Substituting this into \bref{eqn: agh 1}, we derive
\begin{align}
\Gamma a_1
=-\Delta(\Gamma m_2)
=\Gamma(\Delta m_2)
~,
\end{align}
so that $a_1$ turns into the form of  \bref{eqn: trivial agh 1}.
Similarly, when $a_1$ takes the form \bref{eqn: trivial agh 1},
substituting it into \bref{eqn: agh 0}
one sees that $a_0$ takes the form \bref{eqn: trivial agh 0}\,.
We conclude that a  $\Gamma$-trivial $a_2$ leads to a BRST-trivial $S^1$.
So we may set
$
a_2=0
$.
Similarly a $\Gamma$-trivial $a_1$ leads to a BRST-trivial $S^1$.

\section{Derivation of three-, four-, and five-point interaction
of three gauge fields
}\label{app:derivation}

We derive $S^1$, $S^2$ and $S^3$
from \bref{eqn:a1 general}.

\subsubsection*{Three-point interaction}
It is obvious that
$a_1$ in \bref{eqn:a1 general}
satisfies
 $\Gamma a_1=0$ since $\Gamma G(\phib^{(I)})=0$.
Noting that
\begin{align}
-\Delta  a_1=
\intd 
s_{IJ}
G(\phib^{(I)})^T\pat c G(\phib^{(J)})\,,
\end{align}
and that
\begin{align}
&\Gamma\intd \frac{1}{2} s_{IJ}
G(\phib^{(I)})^T\phi G(\phib^{(J)})
\nn\\&
=\intd \frac{1}{2} s_{IJ}G(\phib^{(I)})^{\mu_1\cdots\mu_{s_1}} \pa_{(\mu_1} c_{\cdots \mu_{s_1} \nu_1\cdots \nu_{s_2})} G(\phib^{(J)})^{\nu_1\cdots\nu_{s_2}}
\nn\\&
=\intd 
s_{IJ}G(\phib^{(I)})^T\pat c G(\phib^{(J)})
\,,
\end{align}
we conclude that,
from \bref{eqn: agh 0},
\begin{align}
a_0=\intd \frac{1}{2} s_{IJ}
G(\phib^{(I)})^T\phi G(\phib^{(J)})\,.
\label{eqn:a0 general}
\end{align}
As a result, we obtained
$S^1$ in \bref{eqn:S1}
composed of  \bref{eqn:a2},
\bref{eqn:a1 general}, and \bref{eqn:a0 general}.
\medskip

\subsubsection*{Four-point interaction}
Next we derive $S^2$ in \bref{eqn:S2}.
First, we solve \bref{eqn:ME S2 agh 1} for $b_2$ and $b_1$.
One derives
\begin{align}
-\frac{1}{2}(a_1,a_1)=
\intd\Big[
G(\phib^{(I)})^T\pat c \Gt(\pat c \phibs^{(I)})
 \Big]\,,
\end{align}
as $s_{IJ}s_{JK}=\delta_{IK}$.
Noting that
the right hand side of the above equation
can be expressed as
\begin{align}
\Gamma
\intd G(\phib^{(I)})^T\phi \Gt(\pat c \phibs^{(I)})
+\Delta
\intd \frac{1}{2} \Gt(\pat c \phibs^{(I)})^T\pat c  \phibs^{(I)}\,,
\end{align}
we obtain
\begin{align}
b_2=&
\intd \frac{1}{2} \Gt(\pat c \phibs^{(I)})^T\pat c  \phibs^{(I)}\,,
\label{eqn:b2 general}
\\
b_1=&\intd G(\phib^{(I)})^T\phi \Gt(\pat c \phibs^{(I)})\,.
\label{eqn:b1 general}
\end{align}
It is obvious that \bref{eqn:ME S2 agh 2} is satisfied.
Next, we solve \bref{eqn:ME S2 agh 0} for $b_0$.
It is straightforward to see that
\begin{align}
-\Delta b_1-(a_1,a_0)
=&\Gamma 
\intd
\frac{1}{2}
 G(\phib^{(I)})^T\phi\Gt(\phi G(\phib^{(I)}))\,.
\end{align}
This implies that
\begin{align}
b_0=
\frac{1}{2}
 G(\phib^{(I)})^T\phi\Gt(\phi G(\phib^{(I)}))\,.
 \label{eqn:b0 general}
\end{align}
We conclude that $S^2$  in \bref{eqn:S2} is found to be composed of
\bref{eqn:b2 general}, \bref{eqn:b1 general}, and \bref{eqn:b0 general}.

\subsubsection*{Five-point interaction}

$S^3$ is expanded  as in \bref{eqn:S3}.
First, we solve \bref{eqn:ME S3 agh 2} for $c_2$.
Deriving
\begin{align}
-(a_1,b_2)=
\Gamma \intd
\frac{1}{2}s_{IJ}
\Gt(\pat c \phibs^{(I)})^T\phi \Gt(\pat c\phibs^{(J)})\,,
\end{align}
we obtain
\begin{align}
c_2=\intd \frac{1}{2}s_{IJ}
\Gt(\pat c \phibs^{(I)})^T\phi \Gt(\pat c\phibs^{(J)})\,.
\label{eqn:c2 general}
\end{align}
Next, we examine \bref{eqn:ME S3 agh 1}.
It is easy to derive
\begin{align}
-\Delta c_2=&
\intd s_{IJ}
\Gt(\pat c\phibs^{(I)})^T\phi \Gt(\pat c G(\phib^{(J)}))
\,,
\label{eqn:-Del c2 gene}
\\
-(a_1,b_1)=&
\intd s_{IJ}\Big[
G(\phib^{(I)})^T\pat c \Gt(\phi\Gt(\pat c \phibs^{(J)}))
+\Gt(\phi G(\phib^{(I)}))^T\pat c\Gt(\pat c \phibs^{(J)})
\Big]
\,,
\label{eqn:(a1,b1) gene}
\\
-(a_0,b_2)=&
\intd s_{IJ}
\Gt(\pat c \phibs^{(I)})^T \pat c\Gt(\phi G(\phib^{(J)})) 
\,.
\label{eqn:(a0,b2) gne}
\end{align}
The sum of the right-hand side of \bref{eqn:-Del c2 gene}
and the first term on the right-hand side of \bref{eqn:(a1,b1) gene}
turns into
\begin{align}
\intd s_{IJ} \Gt(\phi(\Gt(\pat c\phibs^{(I)})))^T\pa c G(\phib^{(J)})\,,
\end{align}
while 
 the second term on the right-hand side of \bref{eqn:(a1,b1) gene}
plus
the right-hand side of \bref{eqn:(a0,b2) gne}
becomes
\begin{align}
\Gamma \intd
s_{IJ}
\Gt(\phi G(\phib^{(I)}))^T\phi\Gt(\pat c \phibs^{(J)})
-\intd s_{IJ}\Gt(\pa c G(\phib^{(I)}))^T\phi \Gt(\pat c \phibs^{(J)})
\,.
\end{align}
It follows from these results
that
\begin{align}
-\Delta c_2-(a_1,b_1)-(a_0,b_2)=
\Gamma \intd
s_{IJ}
\Gt(\phi G(\phib^{(I)}))^T\phi\Gt(\pat c \phibs^{(J)})\,,
\end{align}
so that we may obtain
\begin{align}
c_1=\intd
s_{IJ}
\Gt(\phi G(\phib^{(I)}))^T\phi\Gt(\pat c \phibs^{(J)})\,.
\label{eqn:c1 gene}
\end{align}
Finally, \bref{eqn:ME S3 agh 0} is examined.
It is straightforward to see that
\begin{align}
&-\Delta c_1-(a_1,b_0)
\nn\\&=
\Gamma\intd \frac{1}{2}s_{IJ}\Gt(\phi G(\phib^{(I)}))^T\phi\Gt(\phi G(\phib^{(J)}))
-\intd \frac{1}{2}s_{IJ}\Gt(\phi G(\phib^{(I)}))^T \pa c \Gt(\phi G(\phib^{(J)}))
\,,
\label{A gene}
\end{align}
and that
\begin{align}
-(a_0,b_1)=\intd s_{IJ}\Gt(\phi G(\phib^{(I)}))^T \pat c \Gt(\phi G(\phib^{(J)}))
\,
\end{align}
which cancels out the second term on the right-hand side of \bref{A gene}.
As a result, we obtain
\begin{align}
-\Delta c_1-(a_1,b_0)-(a_0,b_1)
=\Gamma\intd s_{IJ} \frac{1}{2}\Gt(\phi G(\phib^{(I)}))^T\phi\Gt(\phi G(\phib^{(J)}))
\,,
\end{align}
which implies that
\begin{align}
c_0=\intd s_{IJ} \frac{1}{2}\Gt(\phi G(\phib^{(I)}))^T\phi\Gt(\phi G(\phib^{(J)}))
\,.
\label{eqn:c0 gene}
\end{align}
Summarizing the results,
we find the $S^3$ in \bref{eqn:S3}
is composed of \bref{eqn:c2 general},
\bref{eqn:c1 gene},
and
\bref{eqn:c0 gene}.


\begin{thebibliography}{9}


\bibitem{Gross}
D.~J.~Gross,
``High-Energy Symmetries of String Theory,''
Phys. Rev. Lett. \textbf{60} (1988), 1229


\bibitem{no-go}
S.~Weinberg,
``Photons and Gravitons in  $S$-Matrix Theory: Derivation of Charge Conservation and Equality of Gravitational and Inertial Mass,''
Phys. Rev. \textbf{135} (1964), B1049-B1056.




\bibitem{KU81}
T.~Kugo and S.~Uehara,
``Massless Particle With Spin $J \geq 1$ Implies the S-Matrix Symmetry''
Prog. Theor. Phys. \textbf{66} (1981), 1044.




\bibitem{Metsaev lc cubic}
R.~R.~Metsaev,
``Cubic interaction vertices of massive and massless higher spin fields,''
Nucl. Phys. B \textbf{759} (2006), 147-201
[arXiv:hep-th/0512342 [hep-th]].

\bibitem{Metsaev lc cubic 2}
R.~R.~Metsaev,
``Cubic interaction vertices for fermionic and bosonic arbitrary spin fields,''
Nucl. Phys. B \textbf{859} (2012), 13-69
[arXiv:0712.3526 [hep-th]].







\bibitem{BBvD85}
F.~A.~Berends, G.~J.~H.~Burgers and H.~van Dam,
``On the Theoretical Problems in Constructing Interactions Involving Higher Spin Massless Particles,''
Nucl. Phys. B \textbf{260} (1985), 295-322.


\bibitem{MMR Noether}
R.~Manvelyan, K.~Mkrtchyan and W.~Ruhl,
``Off-shell construction of some trilinear higher spin gauge field interactions,''
Nucl. Phys. B \textbf{826} (2010), 1-17
[arXiv:0903.0243 [hep-th]];
``A Generating function for the cubic interactions of higher spin fields,''
Phys. Lett. B \textbf{696} (2011), 410-415
[arXiv:1009.1054 [hep-th]].

K.~Mkrtchyan,
``On generating functions of Higher Spin cubic interactions,''
Phys. Atom. Nucl. \textbf{75} (2012), 1264-1267
[arXiv:1101.5643 [hep-th]].


\bibitem{dWitFreedman}
B.~de Wit and D.~Z.~Freedman,
Phys. Rev. D \textbf{21} (1980), 358.


\bibitem{Fronsdal}
C.~Fronsdal,
``Massless Fields with Integer Spin,''
Phys. Rev. D \textbf{18} (1978), 3624.


\bibitem{BRST-antifield}
G.~Barnich and M.~Henneaux,
``Consistent couplings between fields with a gauge freedom and deformations of the master equation,''
Phys. Lett. B \textbf{311} (1993), 123-129
[arXiv:hep-th/9304057 [hep-th]].

M.~Henneaux,
``Consistent interactions between gauge fields: The Cohomological approach,''
Contemp. Math. \textbf{219} (1998), 93-110
[arXiv:hep-th/9712226 [hep-th]].



\bibitem{BRST cohomology}
N.~Boulanger and S.~Leclercq,
``Consistent couplings between spin-2 and spin-3 massless fields,''
JHEP \textbf{11} (2006), 034
[arXiv:hep-th/0609221 [hep-th]].

N.~Boulanger, S.~Leclercq and P.~Sundell,
``On The Uniqueness of Minimal Coupling in Higher-Spin Gauge Theory,''
JHEP \textbf{08} (2008), 056
[arXiv:0805.2764 [hep-th]].








\bibitem{SS2}
M.~Sakaguchi and H.~Suzuki,
in preparation.








\bibitem{AdS/CFT}
J.~M.~Maldacena,
``The Large N limit of superconformal field theories and supergravity,''
Int. J. Theor. Phys. \textbf{38} (1999), 1113-1133,
 Adv. Theor. Math. Phys. 2 (1998) 231-252
[arXiv:hep-th/9711200 [hep-th]].


S.~S.~Gubser, I.~R.~Klebanov and A.~M.~Polyakov,
``Gauge theory correlators from noncritical string theory,''
Phys. Lett. B \textbf{428} (1998), 105-114
[arXiv:hep-th/9802109 [hep-th]].


E.~Witten,
``Anti-de Sitter space and holography,''
Adv. Theor. Math. Phys. \textbf{2} (1998), 253-291
[arXiv:hep-th/9802150 [hep-th]].



\bibitem{3dVasiliev}
S.~F.~Prokushkin and M.~A.~Vasiliev,
``Higher spin gauge interactions for massive matter fields in 3-D AdS space-time,''
Nucl. Phys. B \textbf{545} (1999), 385
[arXiv:hep-th/9806236 [hep-th]];
``3-d higher spin gauge theories with matter,''
[arXiv:hep-th/9812242 [hep-th]].

\bibitem{3dV/W}
M.~R.~Gaberdiel and R.~Gopakumar,
``An AdS$_{3}$ Dual for Minimal Model CFTs,''
Phys. Rev. D \textbf{83} (2011), 066007
[arXiv:1011.2986 [hep-th]].



\bibitem{4dVasiliev}
M.~A.~Vasiliev,
``Nonlinear equations for symmetric massless higher spin fields in (A)dS(d),''
Phys. Lett. B \textbf{567} (2003), 139-151
[arXiv:hep-th/0304049 [hep-th]].



\bibitem{4dV/O}
I.~R.~Klebanov and A.~M.~Polyakov,
``AdS dual of the critical O(N) vector model,''
Phys. Lett. B \textbf{550} (2002), 213-219
[arXiv:hep-th/0210114 [hep-th]].






\bibitem{HLGR fermion BRST-antifield}
M.~Henneaux, G.~Lucena G\'omez and R.~Rahman,
``Higher-Spin Fermionic Gauge Fields and Their Electromagnetic Coupling,''
JHEP \textbf{08} (2012), 093
[arXiv:1206.1048 [hep-th]].


\bibitem{Henneaux:2013gba}
M.~Henneaux, G.~Lucena G\'omez and R.~Rahman,
``Gravitational Interactions of Higher-Spin Fermions,''
JHEP \textbf{01} (2014), 087
[arXiv:1310.5152 [hep-th]].



\bibitem{Sagnotti tensionless limit}
A.~Sagnotti and M.~Taronna,
``String Lessons for Higher-Spin Interactions,''
Nucl. Phys. B \textbf{842} (2011), 299-361
[arXiv:1006.5242 [hep-th]].


M.~Taronna,
``Higher Spins and String Interactions,''
[arXiv:1005.3061 [hep-th]].






\end{thebibliography}
\end{document}